\documentclass[a4paper,twoside]{article}
\usepackage[dvips]{graphicx}
\usepackage{latexsym,epsf}
\usepackage{multicol}
\usepackage[centertags]{amsmath}
\usepackage [cp1251]{inputenc}
\usepackage [ukrainian, russian, english]{babel}
\usepackage {fancyhdr}

\fancyfootoffset{1pt}%
\renewcommand {\footrule}{\vbox to 0pt{\hbox to \headwidth{ \hrulefill \hspace{63mm}}\vss}}

\makeatletter
\renewcommand{\ps@plain}{
\renewcommand{\@oddhead}{}
\renewcommand{\@evenhead}{}
\renewcommand{\@oddfoot}{\hfil \thepage}
\renewcommand{\@evenfoot}{\thepage \hfil \hfil}}
\makeatother 

\newcommand{\be}{\begin{equation}}
\newcommand{\ee}{\end{equation}}

\def\bge{\begin{equation}}
\def\ene{\end{equation}}
\def\bgea{\begin{eqnarray}}
\def\enea{\end{eqnarray}}
\def\nn{\nonumber}
\def\imnim{i}

\def\ls{\raise 1.5pt\hbox{$\,<\;$}\kern -10.5pt\lower3.5pt
          \hbox{$\sim$}\kern 1.5pt} 
\def\gs{\raise 1.5pt\hbox{$\,>\,$}\kern -9.5pt\lower3.5pt
          \hbox{$\sim$}\kern 1.5pt} 

\makeatletter
\renewcommand{\@biblabel}[1]{#1.\hfill}
\makeatother
\title{\textbf{\Large VECTOR TO PSEUDOSCALAR MESON RADIATIVE TRANSITIONS IN CHIRAL THEORY WITH RESONANCES}}
\author{\textbf{\textit{S.A. Ivashyn}}
\footnote{\normalfont
E-mail address: ivashyn@kipt.kharkov.ua}
\\
\emph{\small A.I.~Akhiezer Institute for Theoretical Physics,}\\
\emph{\small National Science Center ``Kharkov Institute of
Physics and Technology'', 61108,  Kharkov, Ukraine}
\\
{\small(Received October 31, 2011)}
}
\begin{document}
\selectlanguage{english}
\date{}
\maketitle

\thispagestyle{fancy}
\begin{center}
\begin{minipage}{165mm}
{\small
%
%
The form factor of the vector meson $V$ 
conversion transition into the pseudoscalar meson $\mathcal{P}$
and the lepton pair $l^+l^-$ is presented
for  $V \equiv$ $\rho^0$, $\omega$, $\phi$ 
 and $\mathcal{P}\equiv$ $\pi^0$, $\eta$.
Our approach is based on the chiral effective 
field theory with resonances.
The normalized form factor for $\omega\to\pi\gamma^\ast$ transition
was fitted to the recent NA~60 data
by varying the only free parameter $\sigma_V$.
The results are compared to the available data
and to the predictions of other models.
 }
\par \vspace{1ex}
\end{minipage}
\end{center}
\begin{multicols}{2}

\begin{center}
\textbf{\textsc{1. INTRODUCTION}}
\end{center}
In current paper we study the conversion transition
of the vector meson $V$ into the pseudoscalar meson $\mathcal{P}$
and the lepton pair $l^+l^-$.
The lepton pair is produced by the virtual photon $\gamma^\ast$:
\bgea
\label{eq:v_decay}
V &\to& \gamma^\ast \mathcal{P} \to l^+l^- \mathcal{P} \ .
\enea
Experimentally, only the processes 
$\omega \to l^+l^-\pi^0$ and $\phi \to l^+l^-\eta$
have been studied out of the whole set of possible combinations
of $V$ and $\mathcal{P}$
($V \equiv$ $\rho^0$, $\omega$, $\phi$ 
 and
 $\mathcal{P}\equiv$ $\pi^0$, $\eta$, $\eta^\prime$).
The most recent information on the former process 
comes from the CERN SPS experiment NA~60~\cite{:2009wb,Usai:2011zza,Uras:2011zz}.
The knowledge of the latter is less precise and given by Novosibirsk
experiment SND~\cite{Achasov:2000ne}.
The measured quantity is the transition form factor 
$\mathcal{F}_{V \to \gamma^\ast \mathcal{P}}(Q^2)$
as a function of the lepton-pair invariant 
mass $Q^2\equiv M_{l^+l^-}\equiv M_{\gamma^\ast}$.

It is known from Refs.~\cite{:2009wb,Usai:2011zza,Uras:2011zz} that
there is no good theoretical description of the recent data 
for $\omega \to l^+l^-\pi^0$.
The most advanced model 
was presented recently in Ref.~\cite{Terschluesen:2010ik},
but still it is not able to accommodate well the full
range of $M_{\gamma^\ast}$.
We would like to stress that a poor knowledge of 
$V \to \gamma^\ast \mathcal{P}$ transition
can be one of the obstacles in the study of light scalar mesons:
it was shown in~\cite{Eidelman:2010ta} that 
the differential cross section of the $e^+e^-$ annihilation
to $\pi^0 \pi^0 \gamma$ and $\pi^0 \eta \gamma$ 
for $\sqrt{s}=M_\phi$ are strongly affected by the
contributions of the type $V \to \gamma^\ast \mathcal{P}$.
The other important observation is that
the $V \to \gamma^\ast \mathcal{P}$ transition 
and scalar meson decay contributions ``contaminate''~\cite{Ivashyn:2009pi}
the pion form factor precision measurement 
with the radiative return method~\cite{Ambrosino:2010bv}.
The above issues make the study of the process~(\ref{eq:v_decay})
interesting.

The first aim of this paper is to present an effective field 
theory description of the processes~(\ref{eq:v_decay}).
We apply the Lagrangian of the chiral perturbation theory with 
resonances~\cite{Ecker:1989yg,Ecker:1988te,Prades:1993ys}.
This effective field theory is a universal tool for 
low-energy particle phenomenology and allows for a consistent
description of the interactions 
of light pseudoscalar mesons ($\pi$, $K$, $\eta$,~\ldots)
with vector ($\rho$, $\omega$, $\phi$,~\ldots),
axial-vector ($a_1$,~\ldots) and
scalar ($a_0$, $f_0$, \ldots) resonances.
The Lagrangian is organized as series in the masses of light quarks 
and derivatives  acting on the pseudoscalar fields,
which are considered as the Nambu-Goldstone fields 
of the spontaneous chiral symmetry breaking.
The expansion coefficients are called the low energy constants (LECs).
The state of the art is that these LECs are obtained from 
the fits to low-energy experimental data.
It is known that this effective field theory is the correct
limit of the Quantum Chromodynamics (QCD) at low energy.
It naturally has the momentum-dependent vertices			   
and exhibits the decoupling in the chiral limit.
The energy scale of the applicability of the chiral effective theory 
with resonances is about $1\;\text{GeV}$.

In chiral theory with resonances, the strength of vector-vector-pseudoscalar meson transition ($VV\mathcal{P}$) is governed by 
the effective coupling $\sigma_V$ which value cannot be theoretically calculated. 
This coupling appears in model description of various processes 
(see, for example,~\cite{Eidelman:2010ta,Ivashyn:2006gf}) and, therefore, 
it is important to estimate its value from experiment. 
However, a direct measurement of this coupling is impossible.

A term proportional to $\sigma_V$ is present in a model description of the vector-pseudoscalar radiative transition form factor for virtual photons~\cite{Eidelman:2010ta}. 
The formulae are given in \textbf{Section~2.}
  The form factors of interest were obtained in the 
  paper~\cite{Eidelman:2010ta}
  only as a by-product and were not compared to data.
In \textbf{Section~3} we demonstrate that the value of $\sigma_V$ coupling can be estimated from fitting the $\omega\to\pi\gamma^\ast$ form factor, recently measured in the NA~60 
experiment~\cite{:2009wb,Usai:2011zza}.
The conclusions are drawn in \textbf{Section~4.}

\begin{center}
\textbf{\textsc{2. FORMALISM}}
\end{center}
We use the chiral Lagrangian in the vector formulation for spin-$1$ fields,
following the papers~\cite{Ecker:1989yg,Ecker:1988te,Prades:1993ys}.
At the moment we neglect the $G$-parity breaking and OZI breaking effects
and use the $SU(3)$ flavor symmetry relations for the couplings.
We also omit the $\eta^\prime$ meson terms for a while.
The Lagrangian terms relevant for the calculation of
$\mathcal{F}_{V \to \gamma^\ast \mathcal{P}}(Q^2)$ 
read
  \begin{equation}
\mathcal{L}_{\gamma V} =  - \, e f_V  \partial^\mu B^\nu \bigl(
\tilde{\rho}^0_{\mu\nu} + \frac{1}{3}\tilde{\omega}_{\mu\nu} -
\frac{\sqrt{2}}{3}\tilde{\phi}_{\mu\nu} \bigr) \ ,
 \label{eq:vector_gamma_V}
\end{equation}
where $\tilde{V}_{\mu \nu} \equiv \partial_\mu V_\nu -
       \partial_\nu V_\mu$;
\begin{eqnarray}
\mathcal{L}_{V\gamma P}&=& -\, \frac{4\sqrt{2} e h_V}{3
f_\pi}\epsilon_{\mu\nu\alpha\beta} \partial^\alpha B^\beta \biggl[ \bigl[
\rho^{0\mu}  +3\omega^\mu \bigr]
\partial^\nu \pi^0\nn \\ \label{lagr_vgp}
&&
 + \bigl[ (3 \rho^{0 \mu} + \omega^\mu)C_q + 2 \phi^\mu C_s \bigr] \partial^\nu \eta
 \biggr] \ ;
\end{eqnarray}
\begin{eqnarray}
\nn
\mathcal{L}_{VVP}&=&-\, \frac{4\sigma_V}{f_\pi}\epsilon_{\mu\nu\alpha\beta}
\biggl[ 
\pi^0 \partial^\mu \omega^\nu \partial^\alpha \rho^{0\beta}
\\
&&
 +  \eta \bigl[  (\partial^\mu\rho^{0\nu} \partial^\alpha
\rho^{0\beta}+
\partial^\mu \omega^{\nu} \partial^\alpha \omega^{\beta} )
  \frac{1}{2}\,C_q
  \nonumber \\
&&
 - \partial^\mu \phi^{\nu}\partial^\alpha \phi^{\beta}
\frac{1}{\sqrt{2}} \, C_s \bigr] \biggr] \ .
\label{lagr_vvp}
\end{eqnarray}
The
$\epsilon_{\mu \nu \alpha \beta}$ is the totally antisymmetric
Levi-Civita tensor.
The pion decay constant is $f_{\pi} = 92.4$~MeV.
The coupling
constants $f_V$, $h_V$ and $\theta_V$ are model parameters:
$f_{V} = 0.20173(86)$ from $\Gamma(\rho^0\to e^+e^-) = \frac{e^4 M_\rho f_{V}^2}{12\pi}$;
$h_{V} = 0.041(3)$ from $\Gamma(\rho\to \pi\gamma) = \frac{4\alpha M_\rho^3 h_{V}^2}{27 f_\pi^2}\left(1-\frac{m_\pi^2}{M_\rho^2}\right)^3$
according to the PDG values for the widths~\cite{Nakamura:2010zzi}.
The coefficients $C_q \approx 0.720$, $C_s \approx 0.471$ account 
for the $\eta$ mixing.

According to the above Lagrangian terms, at the tree level,
the form factors $\mathcal{F}_{V \to \gamma^\ast \mathcal{P}}(Q^2)$ 
read~\cite{Eidelman:2010ta}
\begin{eqnarray}
\label{eq:ff:rho_pi}
\mathcal{F}_{\rho \pi\gamma^\ast } (Q^2)\!\!\! &=& \!\!\!
\frac{4}{3f_\pi} \big[\sqrt{2} h_V - \sigma_V f_V {Q^2} D_\omega (Q^2)
\bigr]
\\
&=& \frac{1}{C_q}\mathcal{F}_{\omega \eta\gamma^\ast }(Q^2)\ ,
\\
\label{eq:ff:omega_pi}
\mathcal{F}_{\omega \pi\gamma^\ast }(Q^2)\!\!\! &=& \!\!\! \frac{4}{f_\pi} 
\bigl[\sqrt{2} h_V - \sigma_V f_V {Q^2} D_\rho (Q^2)
\bigr]
\\
&=& \frac{1}{C_q}\mathcal{F}_{\rho \eta\gamma^\ast }(Q^2)\ ,
\\
\label{eq:ff:phi_eta}
\mathcal{F}_{\phi \eta \gamma^\ast } (Q^2)\!\!\! &=& \!\!\!
\frac{8 C_s}{3f_\pi} \big[\sqrt{2} h_V - \sigma_V f_V {Q^2} D_\phi (Q^2)
\bigr] \ ,
\end{eqnarray}
where the vector meson propagator is
\bgea
D_V(Q^2) \!\!\!\!\!\!  &=& \!\!\!\!\!\!  [Q^2 - M_V^2 + \imnim
\sqrt{Q^2} \Gamma_{tot, V} (Q^2)]^{-1} \, 
\enea
with $M_V = M_\rho, \ M_\omega, \ M_\phi$.

One can expect the $Q^2$-dependent total width 
of the vector meson $\Gamma_{tot, V} (Q^2)$  
to be important only for the $\rho$ meson
within the scope of current research,
because we are interested in the time-like region 
of momenta $0<\sqrt{Q^2}<0.7$~GeV, which partly overlaps 
with the broad $\rho$ resonance.
A possible choice for $\Gamma_{tot, \rho} (Q^2)$ is
(see, for example,~\cite{Eidelman:2010ta})
\bgea
\Gamma_{tot, \rho}(Q^2)\!\! &=& \!\!\frac{G_V^2 M_\rho^2 }{48 \pi f_\pi ^4 Q^2}
\biggl[ \bigl(Q^2 \!-\! 4 m_\pi^2 \bigr)^{3/2}
\theta\bigl(Q^2\! -\! 4 m_\pi^2 \bigr) \nn \\&& \!\! + \frac{1}{2}
\bigl(Q^2 \!- \!4 m_K^2 \bigr)^{3/2}\theta\bigl(Q^2 \!- \!4 m_K^2 \bigr) \biggr]
\ , \label{eq:width:rho}
\enea
where the coupling constant $G_{V}$ 
can be determined from the decay width $\Gamma(\rho^0 \to \pi^+\pi^-)$ 
via 
\bgea
\Gamma(\rho^0 \to\pi\pi) &=& \frac{G_{V}^2}{48\pi f_\pi^4}(M_\rho^2-4m_\pi^2)^{3/2}
 .\label{eq:G_V_1}
\enea
Using the PDG~\cite{Nakamura:2010zzi} value for the width,
one obtains $G_{V} = 65.18(15)$~MeV.

For the $\omega$ and $\phi$ mesons below we use 
the constant width approximation with the width 
values according to PDG~\cite{Nakamura:2010zzi}.

\begin{center}
\textbf{\textsc{3. RESULTS}}
\end{center}
From \textbf{Section~2} we see that the only ambiguous parameter
in the model is $\sigma_V$.
Experimentally, only the normalized form factors are known
\bgea
F_{V \to \gamma^\ast \mathcal{P}}(Q^2) 
&\equiv& 
\frac{\mathcal{F}_{V \to \gamma^\ast \mathcal{P}}(Q^2)}
{\mathcal{F}_{V \to \gamma^\ast \mathcal{P}}(0)} \ .
\enea
We perform the least-square fit of the recent data from CERN SPS NA~60
experiment ($\omega\to\pi\mu^+\mu^-$)~\cite{:2009wb,Usai:2011zza}
by varying the value of $\sigma_V$.
The best $\chi^2$ value is obtained for $\sigma_V\approx 0.58$.
The $\omega \to \gamma^\ast\pi^0$ form factor (normalized)
is shown in~\textbf{Fig.1.}
We observe an unavoidable big discrepancy in the region of 
$M_{\gamma^\ast}>0.6$~GeV, 
however in the rest of the range the theory 
is fairly consistent with the data.

In \textbf{Fig.1.} we can notice that our model 
agrees with the data slightly better than the model 
of Ref.~\cite{Terschluesen:2010ik}, but for both models
the problematic region starts at $M_{\gamma^\ast}\approx0.6$~GeV.
In order to investigate the impact of the $\rho$
meson line-shape on the form factor at $M_{\gamma^\ast}>0.6$~GeV
we perform a complimentary calculation neglecting the resonance width
and make a new fit. It is found that the effects of zero-width 
approximation are almost negligible below $0.6$~GeV 
and are much smaller than the data/model discrepancy above $0.6$~GeV.
A similar conclusion was found also in Ref.~\cite{Terschluesen:2010ik}.
We conclude that the line-shape of the $\rho$ meson alone 
can not be responsible for the discrepancy between the model 
and the data.

It is also interesting to demonstrate that a simple Vector
Meson Dominance ansatz (VMD) 
\bgea
\label{eq:vmd_ff}
F_{V \to \gamma^\ast \mathcal{P}}(Q^2) 
&=&
\frac{M_V^2}{M_V^2 - Q^2}
\enea
utterly fails to describe the data
already as low as at $M_{\gamma^\ast}\approx0.45$~GeV,
see~\textbf{Fig.1.} 
This fact by itself makes the problem of $F_{V \to \gamma^\ast \mathcal{P}}$
modeling very important.

\noindent
\begin{minipage}{80 mm}
\includegraphics[width=\textwidth]{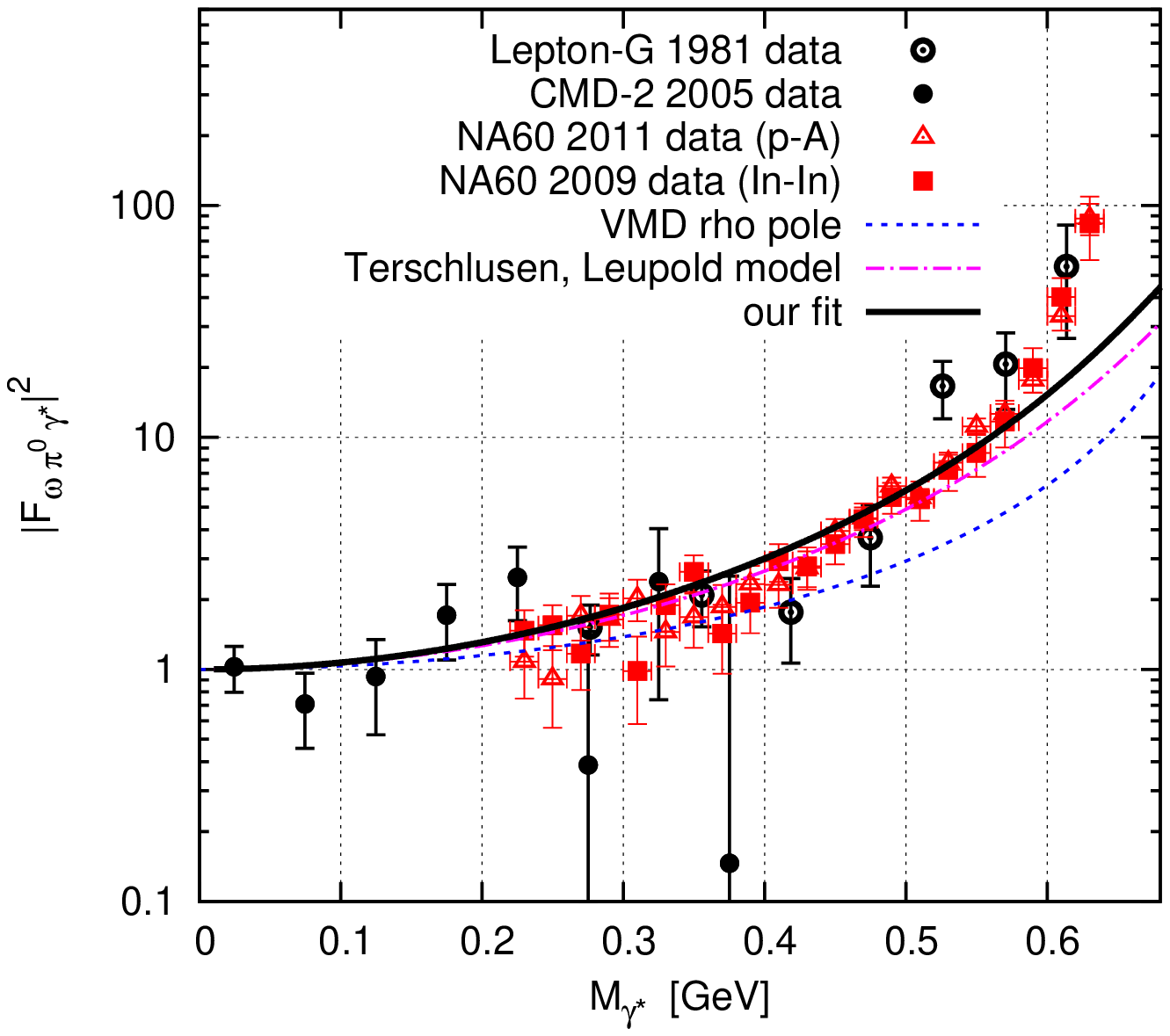}
\emph{\textbf{Fig.1.}} {\emph{
   The $\omega \to \gamma^\ast\pi^0$ form factor \hfill (normalized to unity).
   Lepton-G data is given in Ref.~\cite{Dzhelyadin:1980tj};
   CMD-2 data ---~\cite{Akhmetshin:2005vy};
   NA~60 data --- (p-A)~\cite{Usai:2011zza}, (In-In)~\cite{:2009wb}.
   The model of Terschl\"usen and Leupold is given in Ref.~\cite{Terschluesen:2010ik}.
   VMD rho pole formula --- eq.~(\ref{eq:vmd_ff}) with $M_V=M_\rho$.
}}\\
\end{minipage}

Equations (\ref{eq:ff:rho_pi})--(\ref{eq:ff:phi_eta})
are inter-related via the $SU(3)$ flavor symmetry for the coupling constants
and predict the transition form factors 
for the poorly measured process $\phi\to\eta\gamma^\ast$ 
and for the not measured processes $\rho\to\pi\gamma^\ast$, 
            $\rho\to\eta\gamma^\ast$ and
            $\omega\to\eta\gamma^\ast$.

The model prediction for the 
$\phi \to \gamma^\ast\eta$ form factor (normalized)
is shown in \textbf{Fig.2.}
Here we observe a full consistency with available data from 
Novosibirsk~\cite{Achasov:2000ne}.
We would like to remark that 
new precise data from KLOE experiment will appear soon~\cite{Collaboration:2011zc}
and serve as an important test of the models. 
Obviously, the new data are required in order to cross-check 
the validity of VMD for the case of $\phi \to \gamma^\ast\eta$ transition.

\noindent
\begin{minipage}{80 mm}
\includegraphics[width=\textwidth]{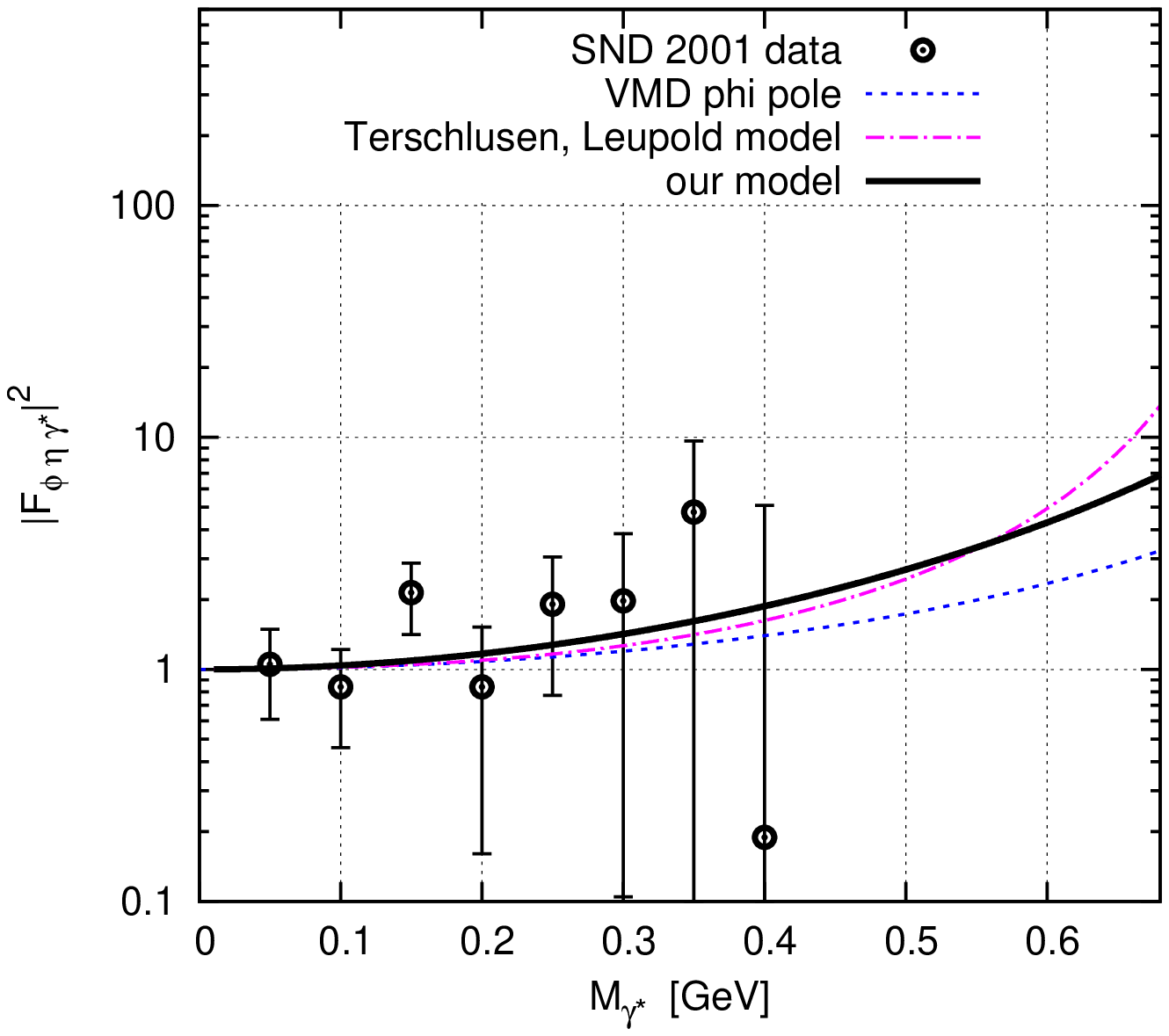}
\emph{\textbf{Fig.2.}} {\emph{
   The $\phi \to \gamma^\ast\eta$ form factor \hfill (normalized to unity).
   SND data is given in Ref.~\cite{Achasov:2000ne}.
   The model of Terschl\"usen and Leupold is given in Ref.~\cite{Terschluesen:2010ik}.
   VMD phi pole formula --- eq.~(\ref{eq:vmd_ff}) with $M_V=M_\phi$.
}}\\
\end{minipage}

\vspace{15pt}

\begin{center}
\textbf{\textsc{6. CONCLUSIONS}}
\end{center}
Within the chiral perturbation theory with resonances,
following the Lagranigian formalism given in 
Refs.~\cite{Ecker:1989yg,Ecker:1988te,Prades:1993ys},
the description of the conversion transition
of the vector meson $V$ into the pseudoscalar meson $\mathcal{P}$
and the lepton pair $l^+l^-$ is presented
($V \equiv$ $\rho^0$, $\omega$, $\phi$ 
 and
 $\mathcal{P}\equiv$ $\pi^0$, $\eta$, $\eta^\prime$).
The normalized form factor for $\omega\to\pi\gamma^\ast$ transition
was fitted to the recent data of CERN SPS experiment 
NA~60~\cite{:2009wb,Usai:2011zza}
and compared to the Lepton-G data~\cite{Dzhelyadin:1980tj};
and CMD-2 data~\cite{Akhmetshin:2005vy}.
It was possible to accommodate the data,
except for the very high $M_{\gamma^\ast}$, 
by adjusting only one model parameter --- $\sigma_V$.
The quality of the data description by our model
is slightly better than that by the model of Ref.~\cite{Terschluesen:2010ik}
and much better than that by the Vector Meson Dominance ansatz.
A possible way to reduce a big data/model 
discrepancy in the region of $M_{\gamma^\ast}>0.6$~GeV
is to go beyond a tree level approximation.
We leave this option for future investigations.

After the value of $\sigma_V$ is fixed, 
the model gives the predictions for various 
$V \to \gamma^\ast \mathcal{P}$
transitions,
among which there are the 
poorly measured process $\phi\to\eta\gamma^\ast$~\cite{Achasov:2000ne}
and the not measured processes $\rho\to\pi\gamma^\ast$, 
            $\rho\to\eta\gamma^\ast$, 
            $\omega\to\eta\gamma^\ast$.
The new precise results on $\phi\to\eta\gamma^\ast$ from 
Frascati~\cite{Collaboration:2011zc}
could help to benchmark the models.

\begin{center}
\textbf{\textsc{7. ACKNOWLEDGEMENTS}}
\end{center}
We profited from discussions with Maurice Benayoun,
Alexandr Korchin, Andrzej Kup\'s\'c and Jaros{\l}aw Zdebik.
We would like to thank Sanja Damjanovic for providing us 
with the data of Ref.~\cite{:2009wb} 
and Gianluca Usai for the data of Ref.~\cite{Usai:2011zza}.
This work is a part of the activity of the ``Working Group on Radiative
Corrections and Monte
Carlo Generators for Low Energies'' {\tt [http://www.lnf.infn.it/wg/sighad/]}.
The partial support from 
Polish Ministry of Science and High Education
from budget for science for years 2010-2013: grant number N~N202~102638
and 
National Academy of Science
of Ukraine under contract $50/53 - 2011$
is acknowledged.

\vspace{3mm}
\begin{center}

\end{center}
\end{multicols}
\selectlanguage{russian} \vspace{8mm}
\begin{center}
\textbf{РАДИАЦИОННЫЕ ПЕРЕХОДЫ ВЕКТОРНОГО МЕЗОНА В ПСЕВДОСКАЛЯРНЫЙ МЕЗОН
В КИРАЛЬНОЙ ТЕОРИИ С РЕЗОНАНСАМИ}\\
\par \vspace{1ex}\textbf{\textit{С.А. Ивашин}}\\
\end{center}
\noindent 
Форм-факторы для конверсионного перехода векторного мезона $V$ 
в псевдоскалярный мезон $\mathcal{P}$
и лептонную пару $l^+l^-$ представлены для
случаев $V \equiv$ $\rho^0$, $\omega$, $\phi$ 
 и $\mathcal{P}\equiv$ $\pi^0$, $\eta$.
Подход основан на эффективной киральной теории поля
с резонансами.
Нормированный форм-фактор для перехода $\omega\to\pi\gamma^\ast$ 
подогнан к данным эксперимента NA~60 	
при помощи варьирования единственного свободного 
параметра $\sigma_V$.
Результаты подхода сравниваются с имеющимися данными
и предсказаниями других моделей.
\\

\selectlanguage{ukrainian}
\begin{center}
\textbf{РАДІАЦІЙНІ ПЕРЕХОДИ ВЕКТОРНОГО МЕЗОНА В ПСЕВДОСКАЛЯРНИЙ МЕЗОН
В КІРАЛЬНІЙ ТЕОРІЇ З РЕЗОНАНСАМИ}\\
\par \vspace{1ex}
\textbf{\textit{С.А. Івашин}}\\
\end{center}
\noindent 
Форм-фактори для конверсійного переходу векторного мезона $V$ 
в псевдоскалярний мезон $\mathcal{P}$
та лептонну пару $l^+l^-$ представлено для
випадків $V \equiv$ $\rho^0$, $\omega$, $\phi$ 
 і $\mathcal{P}\equiv$ $\pi^0$, $\eta$.
Підхід засновано на эффективній кіральній теорії поля
з резонансами.
Нормований форм-фактор для переходу $\omega\to\pi\gamma^\ast$ 
підігнано до даних эксперименту NA~60 
за допомогою варіації єдиного вільного
параметру $\sigma_V$.
Результати підходу порівняно з наявними даними та 
передбаченнями інших моделей.
 \\ \
 \\ \\
\end{document}